\documentstyle[12pt]{article}

\begin{document}

\begin{titlepage}
\begin{flushright}
{\sc SSCL-Preprint-482}\\
{\sc WIS-93/67/July-PH}\\
{\sc CMU-HEP93-10; DOE-ER/40682-35}\\
{\sc \today}\\ [.5in]
\end{flushright}
\begin{center}
{\LARGE Constraints on Extended Technicolor Models
from $B \rightarrow \mu^+ \mu^- X$}\\ [.3in]
{\large Benjam\'{\i}n Grinstein}\\[.1in]
{\small SSC Laboratory, 2550 Beckleymeade Ave.,
Dallas TX 75237} \\[.1in]
{\large Yosef Nir} \\[.1in]
{\small Physics Department, Weizmann Institute of Science, Rehovot 76100,
Israel} \\[.1in]
{\large and}\\ [.1in]
{\large Jo\~{a}o M. Soares}\\ [.1in]
{\small Department of Physics, Carnegie Mellon
University, Pittsburgh, PA 15213} \\[.5in]

{\normalsize\bf Abstract}\\ [.2in]
\end{center}
{\small A recent study by Randall and Sundrum shows that models
of Extended Technicolor (ETC) have interesting implications on rare
$B$ decays. We extend their study to the decay $B\rightarrow\mu^+\mu^-X$.
ETC models with a GIM mechanism predict a decay rate that is a
factor of order 30 above the Standard Model, violating the experimental
upper bound by a factor of 2--4. ``Traditional" ETC models predict
a decay rate that is a factor of order 4 above the Standard Model,
and will be probed when an improvement in the sensitivity of
experiments by a factor of order 2--4 is achieved.}
\end{titlepage}

\section*{}

In a recent paper, Randall and Sundrum \cite{Lisa:93}
studied the contributions to the decays
$b\rightarrow s\gamma$ and $B_s \rightarrow \mu^+ \mu^-$ from various
classes of Extended Technicolor (ETC) models.
Models with and without a GIM mechanism were considered.
In both cases, the radiative decay has roughly the same rate as
in the Standard Model (SM). However, ETC contributions enhance the rate
for $B_s \to \mu ^+\mu ^-$ by one to two
orders of magnitude.  In this paper, we point out that a similar
enhancement occurs for the decay $b \rightarrow s \mu^+ \mu^-$. This is
particularly interesting because there exists an upper bound on this
rate \cite{UA1:91} which is only an order of magnitude above the SM rate
and, furthermore, near term experiments are expected to further improve
this bound \cite{Snowmass}. In this brief note,
we estimate $BR(b\rightarrow s\mu^+\mu^-)$
in the two classes of ETC models discussed in ref.~\cite{Lisa:93},
and compare it to the present experimental upper bound.

The first class
of models considered in ref.~\cite{Lisa:93}, or ``traditional" ETC
models, contains the minimal set of interactions necessary for quark mass
generation. (The study is restricted to models where neither composite
nor fundamental scalars are involved in the quark mass generation.)
It is this very minimal set of interactions that are considered in
the generation of flavor changing neutral currents (FCNC). As such,
it is safe to assume that predictions based on these can be
considered as lower bounds. The caveat, of course, is the possibility of
cancellations once additional interactions are incorporated.  The rates
that we find strongly deviate from the SM ones and, therefore, unless
these additional cancellations are rather precise, one may conceivably
rule out this class of models in the near future.

The second class of models, those with a ``techni-GIM'' mechanism,
incorporate interactions beyond the minimal set of traditional models.
They are theoretically appealing since, unlike traditional ETC,
the massive vector bosons of the ETC interactions
are nearly degenerate, with masses of a few TeV. Although
these are rather light by
traditional ETC standards, this class of models avoids many dangerous low
energy FCNC interactions by incorporating an automatic
techni-GIM cancellation mechanism. In the absence of masses,
the quark sector has a large $SU(3)^3$ flavor symmetry. If the only
parameters that break this symmetry are the quark mass matrices (in the
weak eigenstate basis), all of
the flavor changing interactions must be proportional to them. Techni-GIM
cancellations occur when one rotates to the mass eigenstate basis.
However, as we show below, this class of models seems to violate
the existing bound on the rate for $b \rightarrow s \mu^+ \mu^-$.

The flavor changing interactions that are induced at low energies
by these classes of ETC models, and that are relevant for
$b \rightarrow s \mu^+ \mu^-$, are \cite{Lisa:93}
($i$) the neutral current $Z$-boson coupling
\begin{equation}
\xi^{(i)} \frac{m_t}{16 \pi v}\frac{e}{\cos\theta_W \sin\theta_W}
Y_u^{\dag 33} Y_u^{32} \bar{s}_L \gamma_\mu b_L Z^\mu ,
\label{eq:1}
\end{equation}
which is present and of similar strength in traditional models and in
models with a techni-GIM mechanism, and ($ii$) the 4-fermion interaction
\begin{equation} \xi^{(ii)} \frac{m_t}{4 \pi v^3}Y_u^{\dag 33}
Y_u^{32} \bar{s}_L \gamma_\mu b_L \bar{\mu}_L \gamma^\mu \mu_L,
\label{eq:2}
\end{equation}
which is highly suppressed in traditional models, but not in models with
a techni-GIM mechanism.
Here, $v=246\ {\rm GeV}$ is the electroweak symmetry breaking scale.
The coefficients $\xi^{(i)}$ and $\xi^{(ii)}$ are model-specific. The
normalization of the matrix $Y_u$ is such that $Y_u^{33} =1$.
Following ref. \cite{Lisa:93}, we take for our computations, as
a rough estimate of the couplings, $Y_u^{32} \sim V_{ts}$, $\xi^{(i)}
\sim 1$, and $\xi^{(ii)}\sim 1$ in models with a techni-GIM, while
$\xi^{(ii)}\ll1$ in traditional models. The dominant ETC contributions
to the decay Hamiltonian
\begin{equation}
{\cal H}_{\rm eff}=2\sqrt{2}G_F\sum_{j=8,9}C_j{\cal O}_j, \label{eq:3}
\end{equation}
where
\begin{eqnarray}
{\cal O}_8 &=& V_{ts} (e^2/16\pi ^2)
\bar{s}_L \gamma_\mu b_L \bar{\mu} \gamma^\mu \mu,\nonumber \\
{\cal O}_9 &=& V_{ts} (e^2/16\pi ^2)
\bar{s}_L \gamma_\mu b_L \bar{\mu} \gamma^\mu\gamma_5\mu,  \label{eq:4}
\end{eqnarray}
are then
\begin{eqnarray}
C_8^{\rm (i)} = C_9^{\rm (i)} (1- 4 \sin^2 \theta_W ) &\makebox[.25in]{}&
C_9^{\rm (i)} =\xi^{(i)} m_t / 8 v \alpha_{QED}, \label{eq:5} \\
C_8^{\rm (ii)} = - C_9^{\rm (ii)} &\makebox[.25in]{}&
C_9^{\rm (ii)} =\xi^{(ii)} m_t / 4 v \alpha_{QED}, \label{eq:6}
\end{eqnarray}
for the interactions ($i$) and ($ii$), respectively. There are no
short distance QCD corrections to this effective Hamiltonian. More
precisely, there is no multiplicative correction to $C_8$ since the FCNC
is partially conserved.  There are additive corrections \cite{Grin:89}
to $C_8$ that arise from mixing of the SM four-quark interaction into the
operator ${\cal O}_8$. Since the ETC contributions are much larger than
the SM ones, there are effectively no short distance QCD corrections.

The differential rate $d\Gamma(b \to s \mu^+ \mu^-)/dx$, where $x =
(p_{\mu^+}+p_{\mu^-})^2 /m_B^2$, is usually expressed in units $\Gamma
(b \to c e \bar{\nu}_e)$, in which case the dependence on CKM parameters
drops out. The rate is given by \cite{Grin:89}
\begin{eqnarray}
\lefteqn{\frac{1}{\Gamma(b\to ce\bar{\nu}_e)}d\Gamma(b\to s\mu^+\mu^-)/dx
=}\nonumber \\ & & \frac{\alpha_{QED}^2}{4 \pi^2 f(m_c/m_b)} (1-x)^2
(1+2x)(|C_8|^2 + |C_9|^2)
\end{eqnarray}
($f(0.3)= 0.52$ is a phase space factor)
and is plotted in fig.~1, for $m_t=160\
{\rm GeV}$, for both type ($i$) and ($ii$) contributions. The SM
result \cite{Grin:89} is also given in the figure for comparison.

Similar to the results of ref.~\cite{Lisa:93} for the decay $B_s\to\mu^+
\mu^-$, we find that the rate for $b\to s\mu^+ \mu^-$ is enhanced,
compared to the SM, by a factor of approximately 4 for
type~($i$) interactions, and by a factor of approximately 30
for type~($ii$) interactions. The corresponding integrated rates are given
in table 1 for various values of $m_t$. They should be compared to
the experimental upper bound \cite{UA1:91}:
$BR(B \rightarrow \mu^+ \mu^- X)\leq 5\times10^{-5}$.
We stress that these estimates are very rough since the precise values
of the coefficients $\xi^{(i)}$ and $\xi^{(ii)}$  are not known.
Moreover, the various contributions may interfere destructively.
Fig.~1 shows that destructive interference could significantly reduce
the rate. Nevertheless, it is unlikely
that the resulting cancellations would be so fine-tuned as to
reduce the rate down to a level close to that of the SM. Note that
even the SM contribution could significantly reduce the total rate
through destructive interference with ETC amplitudes.

In ETC models that have a GIM mechanism, the 4-fermion interaction
($ii$) gives the dominant effect in $b\rightarrow s\mu^+\mu^-$.
Our results, given in table 1, imply that this type of ETC models are
ruled out by experiment, unless the $Z$-exchange contribution from ($i$)
adds destructively or if the various couplings of order one are
somewhat suppressed compared to our naive estimates. An
improvement of the experimental bound by a factor of a few would be
sufficient to rule out that case too -- barring delicate cancellations
or fine-tuning. For traditional models, the type ($ii$) interaction is
highly suppressed and the rate is dominated by the type ($i$)
interaction. The rate is lower than in the previous case by an order of
magnitude, and is marginally allowed by the current experimental upper
limit. Again, it should be probed in the near future when on-going
experiments at CLEO and the Tevatron
derive a bound stronger than the present one by a factor of a few.

\section*{Acknowledgements}
This work was initiated at the Workshop on B Physics at Hadron
Accelerators (Snowmass, Colorado, June 1993). The authors are
grateful to the organization for partial financial support.
This work was also partly supported by the U. S. Department of Energy
under Contracts No. DE-FG02-91ER40682 and DE--AC35--89ER40486.
YN is grateful to the theory group at CERN for their hospitality.
YN is an incumbent of the Ruth E. Recu Career Development Chair
and is supported, in part, by the Israel Commission for Basic Research
and by the Minerva Foundation.

\pagebreak
\pagestyle{empty}
\section*{Figure Captions}
\begin{tabbing}
\=Figure 1: \=The differential decay rate $d\Gamma(b\to s\mu^+\mu^-)/dx$
($x = (p_{\mu^+}+p_{\mu^-})^2 /m_B^2$)\\
\>          \>due to the ETC interactions ($i$) and ($ii$) and in the SM,
\\ \>           \>for $m_t = 160\ {\rm GeV}$.\\
\>          \>    \\
\end{tabbing}

\section*{Table Captions}
\begin{tabbing}
\=Table 1: \=The branching ratio for $B \to  \mu^+ \mu^- X$
in the SM, and that due\\
\>         \>to the ETC interactions
($i$) and ($ii$).\\
\>         \>     \\
\end{tabbing}
\vspace{.8in}

\pagebreak
\pagestyle{empty}
\section*{}
\begin{center}
\begin{tabular}{c|ccc}
\multicolumn{4}{c}{{\bf Table 1}}\\ [.2in]
   $m_t$/GeV    &  SM  &  ($i$)  & ($ii$) \\ [.1in] \hline \\
$130$ &  $3.5 \times 10^{-6}$  & $1.5 \times 10^{-5}$  & $1 \times 10^{-4}$
\\ [.1in]
$160$ &  $5.1 \times 10^{-6}$  & $2 \times 10^{-5}$ & $1.5 \times 10^{-4}$
\\ [.1in]
$190$ &  $7.4 \times 10^{-6}$  & $3 \times 10^{-5}$ &  $2 \times 10^{-4}$
\end{tabular}
\end{center}

\end{document}